\documentclass[twocolumn]{aastex631}

\submitjournal{ApJ}
\accepted{October 20, 2023}

\usepackage{hyperref}

\shorttitle{Bjet\_MCMC}

\begin{document}

\title{Bjet\_MCMC: A new tool to automatically fit the broadband SEDs of blazars}

\author[0000-0003-3878-1677]{Olivier Hervet}
\correspondingauthor{Olivier Hervet}
\affiliation{Santa Cruz Institute for Particle Physics and Department of Physics, University of California,\\
   Santa Cruz, CA 95064, USA}
\email{ohervet@ucsc.edu}

\author[0000-0002-0641-7320]{Caitlin A. Johnson}
\altaffiliation{Now at Starry Skies North (\url{https://starryskiesnorth.org})}
\affiliation{Santa Cruz Institute for Particle Physics and Department of Physics, University of California,\\
   Santa Cruz, CA 95064, USA}

\author{Adrian Youngquist}
\affiliation{Department of Mathematics and Statistics, San José State University,\\
San Jose, CA 95192, United States}



\begin{abstract}
Multiwavelength observations are now the norm for studying blazars' various states of activity, classifying them, and determining possible underlying physical processes driving their emission. Broadband emission models became unavoidable tools for testing emission scenarios and setting values to physical quantities such as the magnetic field strength, Doppler factor, or shape of the particle distribution of the emission zone(s). We announce here the first public release of a new tool, \texttt{Bjet\_MCMC}, that can automatically fit broadband spectral energy distributions (SEDs) of blazars. The complete code is available on GitHub and allows testing leptonic synchrotron self-Compton models (SSC), with or without external inverse-Compton processes from the thermal environment of supermassive black holes (accretion disk and broad line region). The code is designed to be user-friendly and computationally efficient. It contains a core written in C++ and a fully parallelized SED fitting method. The original multi-SSC zones model of \texttt{Bjet} is also available on GitHub but is not included in the MCMC fitting process at the moment. We present the features, performance, and results of \texttt{Bjet\_MCMC}, as well as user advice.
\end{abstract}



\section{Introduction: History, features and main results of \texttt{Bjet}}
\label{sec:intro}

The model \texttt{Bjet}, which stands for ``Blob-in-Jet", takes its root in the quick developments of synchrotron-self-Compton (SSC) models in the second half  of the 90's. It corresponds to a period where observational evidence of a compact non-thermal zone flaring in active galactic nuclei (AGN) jets was well established \citep[e.g.][]{Marscher_1985}, general consensus was reached on the AGN unification schemes \citep[e.g.][]{Maraschi_1994, Urry_1995} and the first generation of gamma-ray space telescopes (CGRO) and ground-based very-high-energy atmospheric Cherenkov telescopes (Whipple, CAT) were reaching maturity.
This new generation of telescopes led for the first time to precisely building multiwavelength spectral energy distributions (SEDs) from radio to gamma-rays of the brightest blazars. 

Two main families of synchrotron self-Compton (SSC) models were developed:
\begin{itemize}
    \item One-zone ``pure SSC" models, which were mostly relevant for the blazar subclass of high-frequency synchrotron peaked BL Lacs (HBLs). This class of source  was first defined in 1995 as ``high-energy cutoff BL Lacs," in replacement to previous instrumentally-based classification such as X-ray-- and Radio-selected BL Lacs \citep{Padovani_1995}. We note here that the first SSC theoretical setup can be traced as far back as late 60's \citep{Ginzburg_1965, Ginzburg_1969}.

    \item Models with thermal external inverse-Compton (EIC) from the interaction of high energy particles of the jet with the thermal ambient radiation field surrounding the nucleus due to the accretion disk emission reprocessed by the broad-lines region (BLR). These SSC+EIC models were primarily used in the modeling of flat spectrum radio quasars (FSRQs) such as 3C 279  \citep{Sikora_1994, Ghisellini_1996, Inoue_1996}.
\end{itemize}
In all cases, the high energy emission zone is assumed to be characterized by a compact spherical zone, further referenced as a ``blob", relatively close to the nucleus and moving along the jet at relativistic speed. This blob is isotropically filled with high-energy particles (usually simplified as an electron population) and a tangled magnetic field. The blob radiation in its reference frame is also considered isotropic. Most of the SSC models follow the spherical radiation transfer formula set by \cite{Gould_1979}.
The particle distribution within the blob $\rho(E)$ is characterized by a power-law-like spectrum, with many possible flavors, presenting an average index $\alpha \sim 2-3$ (considering $\rho(E) \propto E^{-\alpha})$. Such a distribution is mostly justified by a process of diffuse shock acceleration \citep[e.g. Section 21.4 in][]{Longair_1994}.

\texttt{Bjet} is part of a second generation of models, called ``multi-zone" \citep[e.g.][]{Ghisellini_2005, Tavecchio_2011}.
A known issue of one-zone models is that they usually poorly picture broadband SEDs below the infrared energy range. It is understood that most of the radio emission of jetted AGN is produced by large emission zones, which can be observed in radio very-long baseline interferometry as pc-to-kpc radio core and radio-knots. The very extended emission ($>100$ kpc) is mostly contributing below the cm wavelengths energy range.

\begin{figure*}[t!]
\begin{center}
	\includegraphics[width= 0.45\textwidth]{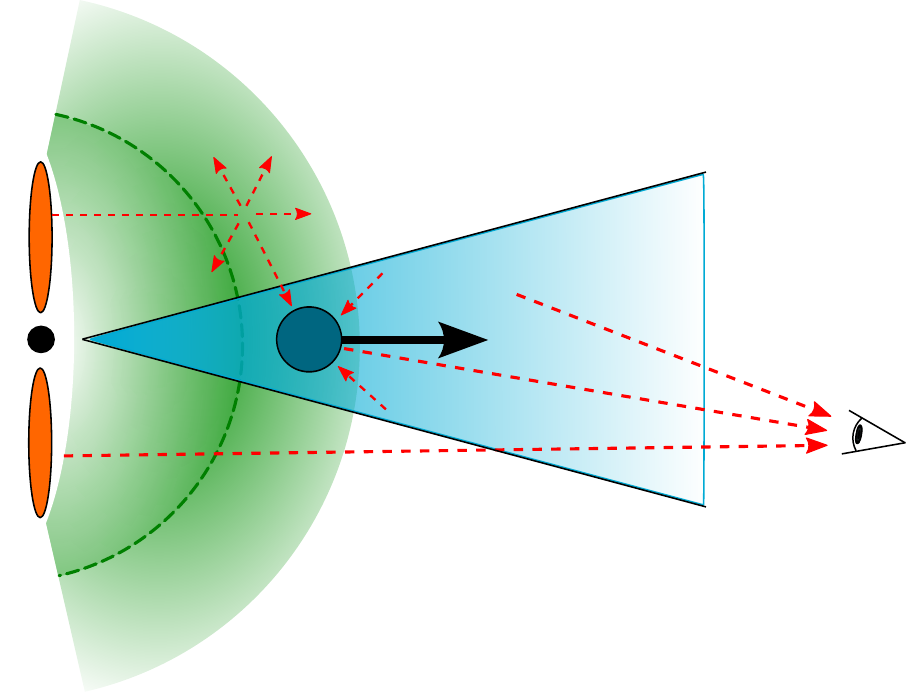}
	\includegraphics[width= 0.52\textwidth]{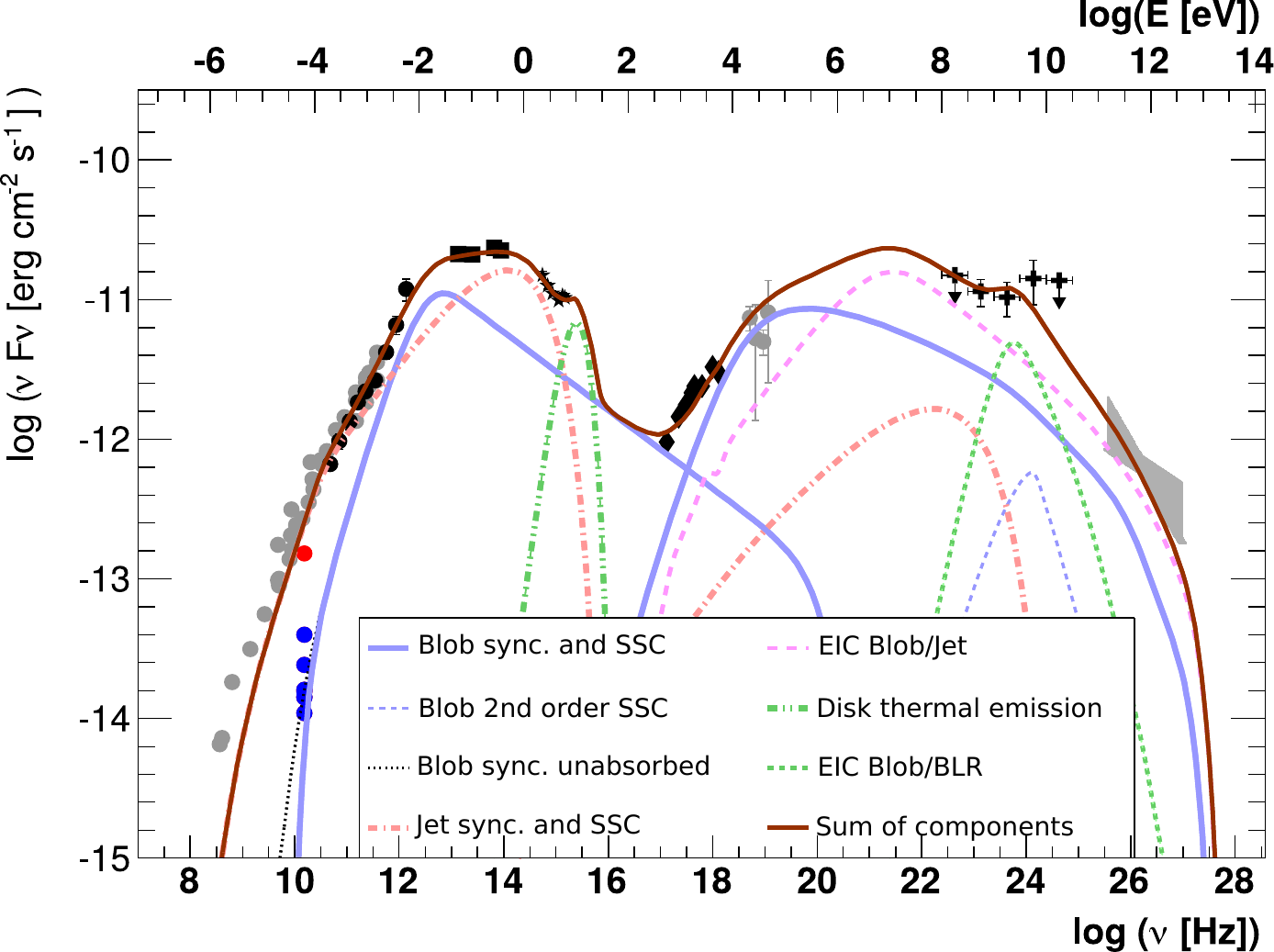}
    \put(-340,105){\colorbox{white}{ \bf Jet}}
    \put(-470,165){\colorbox{white}{ \bf BLR}}
	\put(-530,87){\bf SMBH}
	\put(-535,33){\colorbox{white}{ \bf Accretion disk}}
	\put(-437,67){\colorbox{white}{ \bf Blob}}
	\put(-315,40){\colorbox{white}{ \bf Observer}}
 	\caption{\small{\textit{Left}: Scheme of the \texttt{Bjet} model, dashed lines show the considered radiative transfers. \textit{Right:} Example of an application of the \texttt{Bjet} model on the SED of the blazar AP Librae \citep{Hervet_2015}.}}
 		\label{Fig::Bjet_scheme}
 \end{center}
\end{figure*}
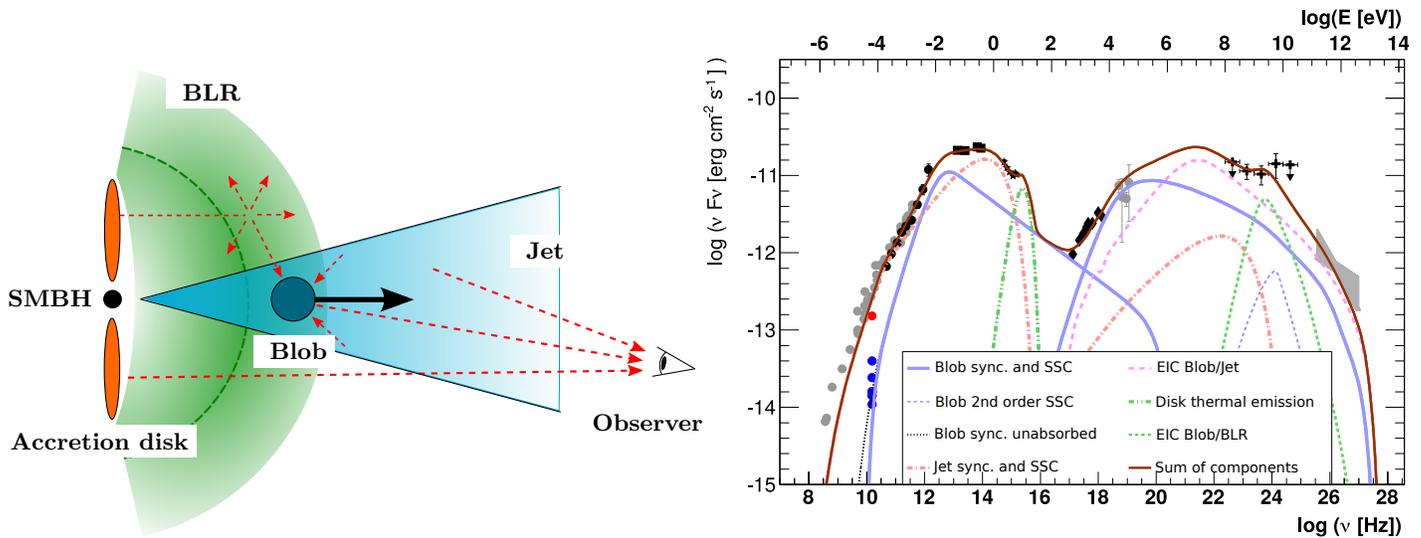

The C++ code foundation that was later used to build \texttt{Bjet} was developed in the early 2000s for a study of the blazar Mrk 501 \citep{Katarzynski_2001}. In order to explain the low-frequency radiation of Mrk 501, they assumed an inhomogeneous model in a conical geometry with a constant bulk Lorentz factor and a power-law decrease of the magnetic field and particle density along the jet. The SSC of an inhomogeneous jet, discretized in homogeneous slices was first developed by \cite{Marscher_1980} and \cite{Ghisellini_1985}.
The approach of \cite{Katarzynski_2001} consisted of two distinct models, one for the blob ``Sblob" and one for the conical jet ``Sjet."
\texttt{Bjet} primary goal was to merge these two models into a consistent multi-zone framework for the study of the  low-frequency-peaked BL Lac (LBL) AP Librae. AP Librae is a blazar that displays a multiwavelength SED with features inconsistent with one-zone models, such as a very broad and relatively flat inverse-Compton emission component from X-rays to very-high-energies (VHE, $E > 100$ GeV). 
This issue was tackled with the self-consistent multi-zone model \texttt{Bjet} that includes radiative interaction between the blob and the conical jet, such as  synchrotron self-absorption, radiative absorption by pair creation, and the external inverse-Compton emission produced by the interaction of the blob's particles onto the jet photons \citep{Hervet_2015}, see Figure \ref{Fig::Bjet_scheme}, \textit{Right}. 

In addition to AP Librae, \texttt{Bjet} has been used for modeling several jetted AGN emitting in VHE, such as PKS 0625–354, HESS J1943+213, 1ES 1215+303, PKS 1222+216 and TON 599 \citep{HESS_2018, Archer_2018, Valverde_2020, Adams_2022}.
It has undergone multiple improvements since its first usage, in optimizing the computation time, but also in its scientific completeness, such as
\begin{itemize}
    \item Better radiation transfer for large angles with the line of sight
    \item External inverse Compton from the blob's particles onto the direct disk radiation
    \item Radial density profile of the broad line region for the thermal EIC and its associated gamma-ray absorption by pair creation. The density profile is based on \cite{Nalewajko_2014}.
\end{itemize}

A geometrical scheme of \texttt{Bjet} (not to scale) is presented in Figure \ref{Fig::Bjet_scheme}, \textit{Left}. In this paper, we do not review the detail of the radiative processes and formula used in the code. They are described in \cite{Katarzynski_2001, Hervet_2015} and, for complementary details, in the Ph.D. thesis \cite[][\textit{in French}]{Hervet_these_2015}.

\section{Motivations for \texttt{Bjet\_MCMC}}

As mentioned above, \texttt{Bjet} has been used in multiple papers since its first development in 2015 and has shown its capability in modeling various types of blazars (HBLs, IBLs, LBLs, FSRQs) and a radiogalaxy candidate (PKS 0625–354). The main purpose of the project \texttt{Bjet\_MCMC} is to provide this tool to the scientific community through an open GitHub project.\footnote{\url{https://github.com/Ohervet/Bjet_MCMC}} 
Users can have access to the full \texttt{Bjet} code and perform one-zone pure SSC, one-zone SSC with thermal nucleus interactions (EIC + pair absorption), and multi-SSC zones with thermal nucleus interactions (blob + jet + nucleus).

The second motivation was to make a tool that automatically fits the multiwavelength SEDs and which is user-friendly and computationally efficient. SSC models are notorious for being challenging for standard $\chi^2$ minimization methods. They have high dimensionalities, parameter degeneracies, local minima, and model-dependant parameter boundaries. To illustrate this last point, let's consider the particle distribution spectrum within the blob, which is set in our model as a broken power law.
\begin{equation}
N_e(\gamma) = \left\lbrace
     \begin{array}{ll}
         N_{e}^{(1)} \gamma^{-n_1} & \mathrm{for}~ \gamma_{\mathrm{min}} \leqslant \gamma \leqslant \gamma_{\mathrm{brk}} \\
         N_{e}^{(2)} \gamma^{-n_2} & \mathrm{for}~ \gamma_{\mathrm{brk}} \leqslant \gamma \leqslant \gamma_{\mathrm{max}}
     \end{array}
     \right. ,
\end{equation}
with $\gamma_{\mathrm{min}}$, $\gamma_{\mathrm{brk}}$ and $\gamma_{\mathrm{max}}$, the Lorentz factors of the radiating particles a the minimum, break, and maximum of their distribution. In this equation, $N_{e}^{(2)} = N_{e}^{(1)}  \gamma_{\rm{brk}}^{(n_2-n_1)}$, and $N_{e}^{(1)}$ the particle density factor set as $N_{e}^{(1)} =N_e(1)$.

Keeping free $\gamma_{\mathrm{min}}$, $\gamma_{\mathrm{brk}}$ and $\gamma_{\mathrm{max}}$ in a standard $\chi^2$ minimization algorithm will certainly create issues since the following condition $\gamma_{\mathrm{min}} < \gamma_{\mathrm{brk}} < \gamma_{\mathrm{max}}$ has to be respected. These constraints let us consider Markov-Chain Monte-Carlo (MCMC) methods as the most suited to perform a SED fit and explore the parameter space. Indeed, MCMCs have the advantage of building the best solution from posterior probability distributions, which is by default less impacted by discontinuity or non-linearity of the parameter space.
We note here that this MCMC fitting approach of SSC models is relatively known in the community and has been implemented and used in multiple studies \citep[e.g.][]{Tramacere_2011, naima_2015, Qin_2018, Jim_2021}

In this paper, we do not intend to describe the general statistical concepts behind the MCMC methods. The literature on the subject is quite vast, we can recommend \cite{MacKay_2003}, for example.

\section{Implementing the MCMC method}

For our project, we used the MCMC \texttt{emcee} package, which is a handy Python tool allowing a relatively simple implementation \citep{Foreman_2013}.\footnote{\url{https://emcee.readthedocs.io/en/stable/}}
\texttt{emcee} requires a user-defined probability function to evaluate the goodness of a fit. It then automatically builds the posterior probability density function following a given number of steps, walkers, and a defined burning sample.
There are multiple flavors available on what type of move a walker can do on the parameter space, we use the default ``StretchMove," developed by \cite{Goodman_2010}. A few other moves -- or combinations of moves -- were tested but did not display significant improvements compared to the proposed default method.

\subsection{Probability function and free parameters}
Our probability function is based on the $\chi^2$ value of the model on all considered SED spectral points. Asymmetric error bars are fully implemented in our $\chi^2$ calculation. We highlight here that flux upper limits are not considered in the fit, but can still be included in the input SED data file for display purposes only. A general advice is to merge constraining upper limits together into larger energy bins until we have a statistically significant data point before fitting the SED. 
As the \texttt{emcee} package requires a log probability, our probability function is defined as $\ln P = -\chi^2 / 2$.

The MCMC method is implemented for the single-zone SSC + thermal EIC model (blob + accretion disk + BLR). It includes up to 13 free parameters, as detailed in the sections below. The full \texttt{Bjet} model currently has 23 free parameters when including the SSC jet. We quickly realized that the computation time required to fit the full multi-zone model is not reasonable for user-friendly usage.\footnote{From rough estimations it would require computation times in the order of months with 10 parallelized cores with the current generation of CPUs.}
In \texttt{Bjet\_MCMC}, the user can decide to fix or free any of the 13 parameters. For pure SSC model fit, the user can deactivate the EIC option to save computation time.

\subsection{Defining the $1\sigma$ parameter space and contour on the SED}
The posterior distribution of probability allows us to define the parameter range corresponding to the $1\sigma$ confidence level ($\sim 68\%$) around the best solution. We follow the general solution proposed by \cite{Lampton_1976, Avni_1976} based on the $\chi^2$ cumulative distribution function $\chi^2_\mathrm{cdf}$, or more precisely on the percent point function $\chi^2_\mathrm{ppf}$ which returns the $\chi^2$ value associated with a probability $P$ and a number of degrees of freedom $k$ of the $\chi^2_\mathrm{cdf}$.
In this approach, we consider all models which have $\chi^2 < \chi^2_\mathrm{min} + \Delta \chi^2$ as within $1\sigma$ of the best value, where $\chi^2_\mathrm{min}$ is our best solution and $\Delta \chi^2 = \chi^2_\mathrm{ppf}(0.682, k)$. $k$ is the number of free parameters in our model that can range from 1 to 13.

\texttt{Bjet\_MCMC} draws the $1\sigma$ contour in the SED associated with the parameter uncertainties. Solutions to get the exact contours were ruled out as too computing and disk-space demanding. For example, one can save all SED points for all models tested during the MCMC process, or run through thousands of randomly picked models within the $1\sigma$ parameter space. In order to save computation time, this contour is built by picking up models with extremum parameter values at the $1\sigma$ confidence level. We call it the ``min-max method." This allows us to build relatively good contours by re-running the code \texttt{Bjet} only twice the number of free parameters, which typically takes up to a few minutes with all 13 parameters free. Hence we must warn the user that the contours on the SED plots are an approximation and do not extend to the full theoretical space covered by the parameters uncertainties.

\subsection{Parameters boundaries}

The pure SSC model has 9 free parameters by default, in addition to 3 fixed parameters:
\begin{itemize}
    \item The redshift $z$ is set by the user.
    \item The cosmology is set by default as a ﬂat $\Lambda$CDM with H0 = 69.6 km s$^{-1}$ Mpc$^{-1}$, $\Omega M$ = 0.286,
and $\Omega \Lambda$ = 0.714 \citep{Bennett_2014}.
    \item The angle between the jet direction and the observer line of sight $\theta$ is fixed at 0.57 degrees to satisfy the Doppler boosted regime $\delta \sin{\theta} < 1$, with the Doppler factor $\delta \leqslant 100$.
\end{itemize}

\begin{table}[h!]
    \centering
    \caption{Parameters and bounds.}
    \begin{tabular}{llll}
    \hline
     \textbf{}   &  \textbf{Range} & \textbf{Description} & \textbf{Scale} \\
    \hline
    \multicolumn{4}{c}{SSC Blob } \\
    \hline
    $\delta$	&	[1 , 100]$^+$	 &      Doppler factor         &         linear\\
    $N_{e}^{(1)}$        &   [0 , 8]	 &      particle density factor [cm$^{-3}$]     &   log10\\
    $n_1$	        &   [1 , 5]$^+$	 &      first index    &      linear\\
    $n_2$	        &   [1.5 , 7.5]$^+$	 &      second index    &      linear\\
    $\gamma_{\mathrm{min}}$	        &   [0 , 5]$^+$	 &      low-energy cutoff    &      log10\\
    $\gamma_{\mathrm{max}}$	        &   [3 , 8]$^+$	 &      high-energy cutoff    &      log10\\
    $\gamma_{\mathrm{break}}$	    &   [2 , 7]$^+$	 &      energy break    &      log10\\
    $B$	    &   [-4 , 0]	 &      magnetic field strength [G]    &      log10\\
    $R$	    &   [14 , 19]$^+$	 &      blob radius [cm]    &     log10\\
    \hline
    \multicolumn{4}{c}{Additional parameters for thermal EIC} \\
    \hline
    $T_\mathrm{disk}$	    &   [3.5 , 6]	 &      disk temperature [K]    &     log10\\
    $L_\mathrm{disk}$	    &   [40 , 50]	 &      disk luminosity [erg/s]    &     log10\\
    $\epsilon_\mathrm{BLR}$	    &   [-5 , 0]	 &  covering factor of the BLR    &     log10\\
    $D_\mathrm{BH}$*    &   [15 , 21]$^+$	 &  Distance blob - SMBH [cm]    &     log10\\
    \hline
    \multicolumn{4}{l}{\footnotesize{* Host galaxy frame.}}\\
    \multicolumn{4}{l}{\footnotesize{$^+$ Parameters with additional constraints.}}
     \end{tabular}
    \label{tab:param_bounds}
    \vspace{0.5cm}
\end{table}

 In the MCMC implementation, we set minimum and maximum values for each of the parameters, as shown in Table \ref{tab:param_bounds}. We intentionally use wide ranges for parameters to make as few assumptions as possible. Many parameters are in log scale to ease walkers' moves through multiple orders of magnitude. 

 \begin{figure*}[t]
\centering
\includegraphics[width=.7\textwidth]{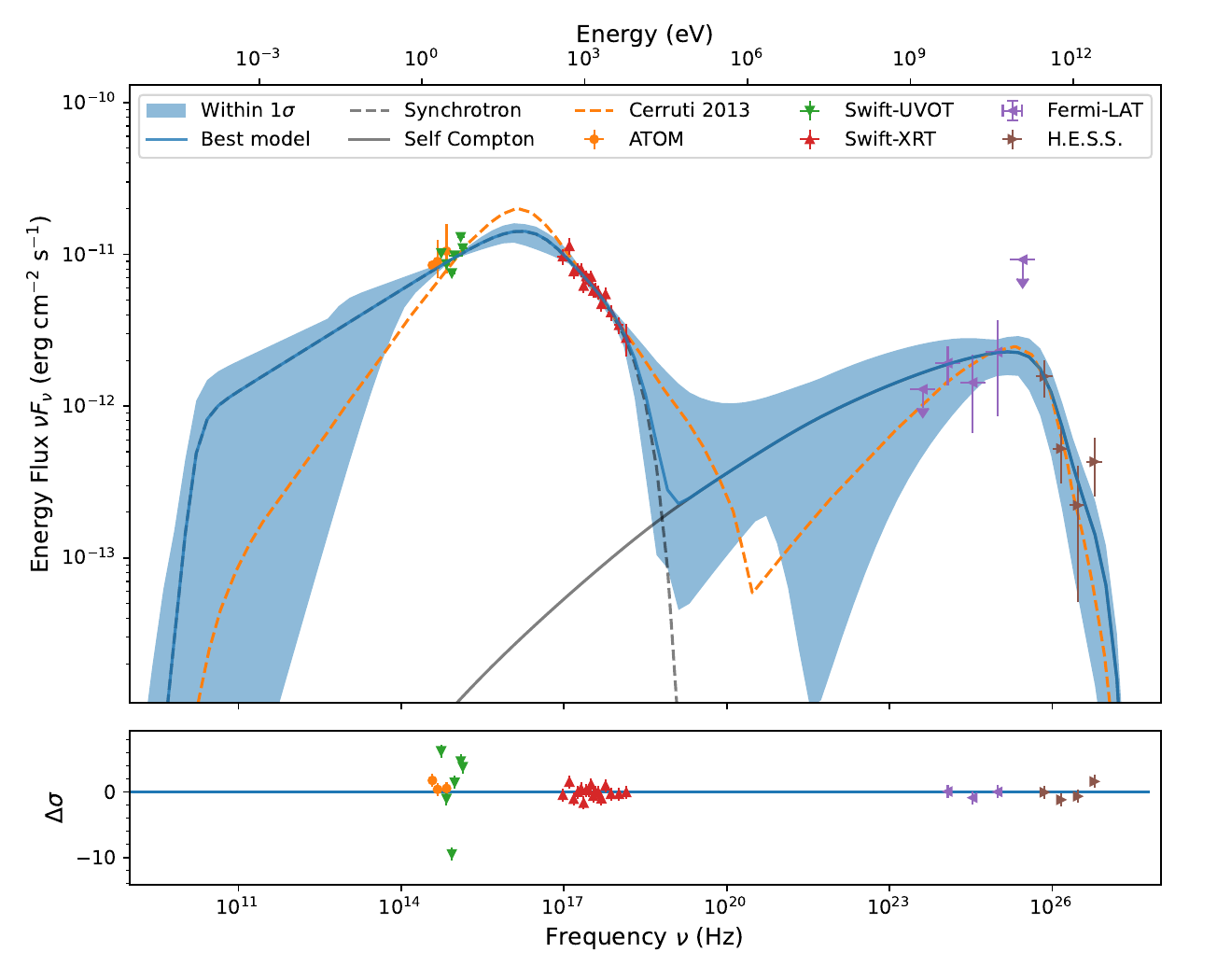}
\caption{Results of \texttt{Bjet\_MCMC} on the SED fit of the blazar 1RXS J101015.9-311909. The model applied was a one-zone pure SSC, including a gamma-ray EBL absorption following the model of \cite{Franceschini_2017}. The SED data points have been shared by the authors of \cite{Cerruti_2013}, while their model line itself has been manually digitized from the original paper.}
\label{Fig:J1010_SED}
\end{figure*}

Within our MCMC method, parameter boundaries mean that the posterior probability $\ln{P} = 0$ for the fit with any parameter outside the given parameter range. The MCMC algorithm acts as an acceptance/reject method only based on the change of the posterior probability value from one walker step to the other. If a walker moves outside the parameter range, the move will be automatically rejected and the walker will try again.
From the \texttt{emcee} package, a good acceptance rate is about 0.2. Given the additional parameter constraints developed below,  \texttt{Bjet\_MCMC} shows an acceptance rate in the order of $\sim 0.05 - 0.1$ from previous tests. 

As highlighted in Table \ref{tab:param_bounds}, multiple parameters have fluctuating boundaries intertwined with other parameter values. The particle spectrum, for example, must follow the condition of $\gamma_{\mathrm{min}} < \gamma_{\mathrm{brk}} < \gamma_{\mathrm{max}}$, and $n_2 > n_1$.
The fastest observed variability $\Delta t_{\mathrm{obs,min}}$ is also used to constrain the Doppler factor and size of the emitting region. From the simple argument that in the jet frame, the fastest variation cannot happen faster than the time the light takes to cross the blob radius, we apply the condition $R \leq {c \delta \Delta t_{\mathrm{obs,min}}}/(1 + z)$. Finally, a last condition is applied, specifying that the blob diameter cannot be larger than the jet cross-section.
From a radio study of a large sample of blazars, it can be noted that the intrinsic jet half-opening angle $\alpha_{\mathrm{jet }/2}$ is no more than 5 degrees. \citep[e.g.][]{Hervet_2016}. Using this conical jet approximation we set the condition $ \alpha_{\mathrm{jet }/2} = \arctan(R/D_\mathrm{BH}) 180/\pi < 5$.

\section{Pure SSC validation on the HBL 1RXS J101015.9-311909}

\begin{table*}[t!]
\begin{center}
\caption{\textbf{Best fit and parameter comparison for 1RXS J101015.9-311909}}
\label{tab:cerruticompare}
\begin{tabular}{l|cc|cc}
     & \multicolumn{2}{c}{Cerruti et al. 2013} & \multicolumn{2}{c}{\texttt{Bjet\_MCMC}} \\
 Parameter & Best Value & $1\sigma$ Range & Best Value & $1\sigma$ Range \\
 \hline
 $\delta$ & 96.83 & 32.07--99.53 & 83.8 & 35.8 -- 100 \\
 $N_{e}^{(1)}$ [cm$^{-3}$] & undefined & undefined & 4.24$\times 10^3$ & 2.18$\times 10^2$ -- $1.0\times 10^6$ \\
 $n_1$ & 2.0 & fixed & 2.56 & 2.31 -- 2.74 \\
 $n_2$ & 4.0 & fixed & 3.75 & 3.25 -- 4.24 \\
 $\gamma_{min}$ & 100 & fixed & 9.22 & 1.00 -- $1.49\times 10^4$ \\
 $\gamma_{max}$ & 5$\times 10^{6}$ & fixed & 1.99$\times 10^6$ & 2.57$\times 10^5$ -- 9.99$\times 10^7$ \\
 $\gamma_{break}$ & 5.31$\times 10^{4}$ & (3.48--13.15)$\times 10^{4}$ & 2.13$\times 10^5$ & 2.40$\times 10^4$ -- 3.93$\times 10^5$ \\
 $B$ [G] & 0.015 & (0.51 -- 4.089)$\times 10^{-2}$ & 1.71$\times 10^{-3}$ & 7.92$\times 10^{-4}$ -- 1.42$\times 10^{-1}$ \\
 $R$ [cm] & 1.3$\times 10^{16}$ & (0.49--11.57)$\times 10^{16}$ & 1.40$\times 10^{17}$ & 1.71$\times 10^{15}$ -- 2.21$\times 10^{17}$ \\
 \hline
 $\chi^2_\mathrm{red}$ total & \multicolumn{2}{c}{260.6/27 = 9.65} & \multicolumn{2}{c}{187.1/24= 7.80}\\
 $\chi^2_\mathrm{red}$ X-ray -- $\gamma$-ray & \multicolumn{2}{c}{18.6/18 = 1.04} & \multicolumn{2}{c}{15.0/15= 1.00}\\
\end{tabular}
\end{center}
\end{table*}

The blazar 1RXS J101015.9-311900 is a high-frequency peaked BL Lac (HBL) with a redshift of $z = 0.143$ that has been discovered emitting up to a few TeV by the H.E.S.S. Collaboration after an observing campaign between 2006 and 2010 \citep{HESS_2012}.
The multiwavelength SED of this source has been successfully fitted with a one-zone SSC model by \cite{Cerruti_2013}. In their study, they developed a fitting algorithm that relies on a strong parametrization of the SED features such as slopes and peaks, and extended the approach made by \cite{Tavecchio_1998}. They also probed the parameter space of their SSC model by  discretizing it in a grid, each free parameter is divided into 10 points, with the exception of the particle index $n_1$ in only 3 points. This kind of probing method quickly becomes very computationally heavy for large dimensionality and has only 6 parameters that can be considered free to mitigate the computation time. On these parameters, they used a different definition of the blob density, which does not allow a straightforward comparison with our results.


Nevertheless, this multiwavelength SED is ideal for a validation test of \texttt{Bjet\_MCM} and allows a partial comparison of results with the work of \cite{Cerruti_2013}. For this model, we considered 100 walkers, 5000 steps, and a burning phase of 200 steps. We ran it over 15 parallelized threads over a full time of 5h 45min.
First, we can visually compare the two models on Figure \ref{Fig:J1010_SED}. We see notable differences but a relatively equally good fit at first glance. We however achieved a better $\chi^2_\mathrm{red}$ considering the full dataset, or only the X-ray to gamma-ray dataset as used by \cite{Cerruti_2013}. These values are reported in Table \ref{tab:cerruticompare}.
We observe a good agreement within errors on our free parameters. A notable difference is that the best value found of $n_1$ with \texttt{Bjet\_MCMC} is outside the probed parameter range of \cite{Cerruti_2013}. 
Overall this comparison fully validates our approach for single-zone pure SSC models by providing to date the best SED fit and parameter characterization on the blazar 1RXS J101015.9-311909.

\section{Fitting the FSRQ PKS 1222+216 with SSC + thermal EIC}

\begin{figure*}[t!]
\centering
\includegraphics[width=0.7\textwidth]{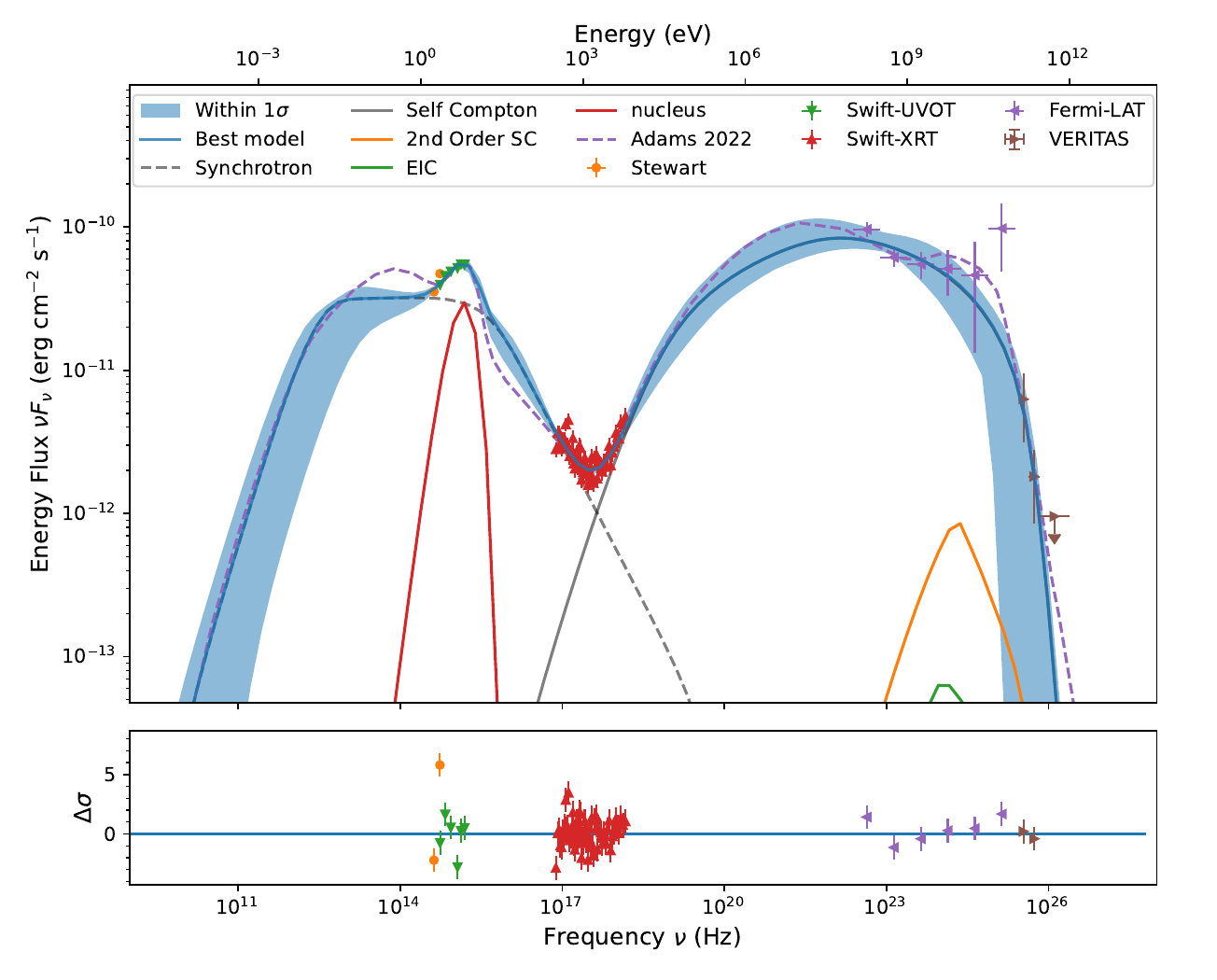}
\caption{Results of \texttt{Bjet\_MCMC} on the SED fit of the blazar PKS 1222+216. The model applied was a one-zone SSC +  thermal EIC from the disk and BLR radiative interaction, including a gamma-ray EBL absorption by the model of \cite{Franceschini_2017}. The SED data points have been shared by the authors of \cite{Adams_2022}, while their model line itself has been manually digitized from the original paper.}
\label{Fig:PKS1222_SED}
\end{figure*}

\begin{table*}[t!]
\begin{center}
\caption{\textbf{Best fit and parameter comparison for PKS 1222+216}}
\label{tab:cerruticompare}
\begin{tabular}{l|c|cc}
     & \multicolumn{1}{c}{Adams et al. 2022} & \multicolumn{2}{c}{\texttt{Bjet\_MCMC}} \\
 Parameter & Best Value  & Best Value & $1\sigma$ Range \\
 \hline
 $\delta$ & 40 & 31.5 & 26.0 -- 42.3 \\
 $N_{e}^{(1)}$ [cm$^{-3}$] & 2.0$\times 10^4$ & 1.12$\times 10^7$ & 3.41$\times 10^5$ -- $9.97\times 10^7$ \\
 $n_1$ & 2.1 & 2.98 & 2.51 -- 3.12 \\
 $n_2$ & 3.9 & 4.45 & 4.14 -- 4.83 \\
 $\gamma_{min}$ & $5.5\times 10^2$ & $7.94\times 10^2$ & $(3.43 - 9.51)\times 10^2$ \\
 $\gamma_{max}$ & 3.0$\times 10^{5}$ & 5.15$\times 10^6$ & 9.42$\times 10^4$ -- 9.96$\times 10^8$ \\
 $\gamma_{break}$ & 5.0$\times 10^{3}$ & 3.09$\times 10^4$ & 4.40$\times 10^3$ - 5.22$\times 10^4$ \\
 $B$ [G] & 3.0$\times 10^{-2}$ & 3.92$\times 10^{-2}$ & 1.25$\times 10^{-2}$ -- 8.06$\times 10^{-1}$ \\ 
 $R$ [cm] & 5.5$\times 10^{16}$ & 7.10$\times 10^{16}$ & 1.94$\times 10^{15}$ -- 1.99$\times 10^{17}$ \\
 $T_\mathrm{disk}$ [K] & 2.8$\times 10^{4}$ & 2.7$\times 10^{4}$ & $(2.23 - 3.35)\times 10^{4}$ \\
 $L_\mathrm{disk}$ [erg s$^{-1}$] & 2.8$\times 10^{46}$ & 2.8$\times 10^{46}$ & fixed \\
 $\epsilon_\mathrm{BLR}$  & 2.0$\times 10^{-2}$* & 2.0$\times 10^{-2}$ & fixed \\
 $D_\mathrm{BH}$ [cm] & 1.10$\times 10^{19}$ & 4.56$\times 10^{19}$ & $2.96\times 10^{18} - 1.00\times 10^{21}$ \\
\hline
 $\chi^2_\mathrm{red}$ & \multicolumn{1}{c}{197.5/66 = 3.04} & \multicolumn{2}{c}{138.6/67 = 2.07}\\
\end{tabular}
\tablenotetext{*}{\textit{The value $\epsilon_\mathrm{BLR}$ was considered as fixed in the study of \cite{Adams_2022}.}}
\end{center}
\end{table*}

In order to further display the capabilities of \texttt{Bjet\_MCMC}, we performed a test on the FSRQ PKS 1222+216 (z = 0.432), which is known to display bright gamma-ray outbursts \citep[e.g.][]{Tavecchio_2011, Adams_2022}. It has been noticed that this blazar was a good candidate for SSC + thermal EIC models as EIC was proposed as the main source of VHE gamma-rays.
However, without a proper fitting method, it is challenging to fully discard the one-zone pure SSC. This point may actually be the most critical in the relevance of \texttt{Bjet\_MCMC} as it is to date likely the first public code that can provide a fit with a full SSC+EIC model (13 free parameters). As pure SSC and SSC+EIC models are nested, \texttt{Bjet\_MCMC} can provide a statistical test that allows to reject the pure SSC hypothesis.

It has to be noted that  MCMC methods can still struggle in complicated parameter spaces and be stuck in local minima. Eventually, a proper MCMC algorithm should always end up in the real best solution. The computation time needed can however be unachievable with standard computers. 

In this section, we compare the results of \texttt{Bjet\_MCMC} on the 2014 flare SED of PKS 1222+216 published by \cite{Adams_2022}. In their paper, the SED model was manually crafted through a ``fit by eye" model using \texttt{Bjet}. Being fitted with the same core code (with only minor updates since then), we can have a proper parameter-by-parameter check on how the results of \texttt{Bjet\_MCMC} differ from the previous model.

For this fit, we used the same MCMC setup as for 1RXS J101015.9-311900 (100 walkers, 5000 steps, 200 steps of burning phase, 15 computing threads). Activating the interaction with the thermal nucleus emission significantly increases the computation time for each step. The full MCMC chain took a total of 12h to run.
After noticing multiple $\chi^2$ local minima, we fixed two parameters to the \cite{Adams_2022} value which are the BLR covering factor $\epsilon_\mathrm{BLR} = 1\times 10^{-2}$ and the accretion disk luminosity $L_\mathrm{disk} = 2.8\times 10^{46}$ erg s$^{-1}$.
As seen in Figure \ref{Fig:PKS1222_SED}, the best fit is visually convincing. We also observe in Figure \ref{Fig::Chi2_PKS1222} that the MCMC walkers display a good general $\chi^2$ convergence. However, we still have hints of local minima, such as in Figure \ref{Fig::Chi2_PKS1222}, \textit{Top-left},  where a small fraction of walkers get stuck away from the best fit. They also appear as ``islands" in the parameter space corner plot (see Figure \ref{Fig::Corner_PKS1222}).

Results of \texttt{Bjet\_MCMC} show better fit $\chi^2$ compared to the model of \cite{Adams_2022}. It is interesting to notice that the best solution of \texttt{Bjet\_MCMC} does not favor any significant contribution for EIC emission in gamma rays. However, it does not rule out a strong EIC emission either. 
In the study of \cite{Adams_2022}, it was estimated that the distance to the SMBH should be at least about one parsec to avoid too much gamma-gamma absorption from the BLR. This is relatively consistent with the estimated distance of $D_\mathrm{BH}$ from our fit which is found at above 0.96 pc from the SMBH.

\section{Computational performances and general using advice}

\texttt{Bjet\_MCMC} makes full use of the paralellized capability of the \texttt{emcee} package. By running several tests, we observed a roughly linear improvement in the computation time following the number of parallel threads used for the fitting process. We have not performed extensive testing to check if this linear relation was holding true at more than about a dozen of threads. It is expected that I/O processes will diminish the relevance of large parallelization at some point. We recommend using a large number of computing threads if available, likely at least 4 for the pure SSC and at least 15 for SSC+EIC if a user wants to get results overnight. \texttt{Bjet\_MCMC} will be the most relevant if used in a computing center with several tens of available computing threads.

\subsection{Quality checks and length of MCMC chains}
We propose a few ways to estimate if a user gets enough walkers and steps to be confident in the output of \texttt{Bjet\_MCMC}, with some warnings and advice.
The favored test to check if the fit is optimal is to get a look at the ``average $\chi^2$ per step" plot. For example, one can see in Figure \ref{Fig::Chi2_1RXS} for 1RXS J101015.9-311909 that the average $\chi^2$ plateaued at about 2500 steps. We can confidently deduce that only 3000 steps would have been enough for this fit as no further improvements are observed afterward. Now looking at the same plot from PKS 1222+216 in Figure \ref{Fig::Chi2_PKS1222}, we observe a good convergence of the average $\chi^2$ but not a full plateauing yet. This means that the full extent of the 1-sigma parameter space is likely going to change marginally. This is usually not a big issue, but one should avoid drawing too firm a conclusion from the exact number of the error associated with parameters. A good practice would be to add an extra 20\% on the parameter errors to get a more conservative parameter range when the average $\chi^2$ curve does not fully flatten out. If the average $\chi^2$ curve does not show any sign of asymptotical behavior, then the number of steps and/or walkers needs to be increased.

Note that the best $\chi^2$ always converges faster than the average $\chi^2$ (see Figures \ref{Fig::Chi2_1RXS}, \ref{Fig::Chi2_PKS1222}). The best $\chi^2$ convergence gives a confidence estimation on the best model while the average $\chi^2$ convergence gives a confidence estimation of the associated parameter errors. The plot ``$\chi^2$ per step" gives a view of the entirety of walkers. It is a relatively efficient way of checking for local $\chi^2$ minima in the parameter space. As soon as most walkers converge toward the best solution, this should not have significant consequences in the results. If most of the walkers appear to be stuck in local minima, then your SED dataset is not constraining enough for the complexity of the model. You can run a longer chain to hope for the MCMC method to eventually converge, or reduce the model complexity by freezing parameters.

\texttt{emcee} considers unsafe a number of steps fewer than 50 times the integrated autocorrelation time $\tau_\mathrm{corr}$ (in steps). We find this boundary challenging to achieve in most of our tests, but also not systematically leading to better results when met. The value $\tau_\mathrm{corr}$ is given as an output of \texttt{Bjet\_MCMC}. We generally advise a safe boundary of $n_\mathrm{steps} \geq 10 \tau_\mathrm{corr}$. However, a general check on the $\chi^2$ convergence plots and a visual check on the SED itself should always be given to assess the fit quality.

\subsection{Advice in building a MWL SED}

Blazars showing great variability should be treated with caution when building their MWL SED. \texttt{Bjet\_MCMC} is a stationary model, which means that within the period considered for the SED, we assume an equilibrium between particle injection and cooling mechanisms. The code cannot model any flux variation or describe any time lags between flares observed at different energies. 
It is possible to model the SED of a flare with \texttt{Bjet\_MCMC}, such as PKS 1222+216 in Fig. \ref{Fig:PKS1222_SED}, but we need to ensure that the time period used for modeling can be considered as steady. It is done, for example, by using a time period much smaller than the duration of the flare. Longer integration times are often used for lower-sensitivity instruments, and non-simultaneous datasets are often used to build a more detailed SED. There is no single way of building a MWL SED since most of the time, it consists of finding the best trade-off between data quantity and quality.
In any case, datasets, time periods, and assumptions used to build SEDs should be documented by users of \texttt{Bjet\_MCMC}. The modeling results are as relevant as the prior assumptions used to build a SED.

Some spectral points with very small errors may be overconstraining the fit to the detriment of another energy range. For example, a 10\% model flux variation in optical may be as constraining as a factor 2 for gamma rays. From a statistical point of view, this is exactly how we expect the fit to behave following the error bars of spectral points. However, the flux ``real" errors are often widely underestimated in low energies. 
The error in each spectral point should reflect the flux variations during the integration period used to build the SED. We advise users to use the RMS error of flux variation instead of the statistical error of individual observations, when larger. It appears that both optical SEDs used in this paper likely have overconstraining error bars.

\subsection{Other considerations}
A full version of the multi-zone \texttt{Bjet} code is available to be used through the package \texttt{Bjet\_MCMC}, but contains a full amount of 23 free parameters, and consequently is not included in the MCMC method. The multi-zone model is recommended to be used only by scientists who have a deep knowledge of blazar emission models as there are significant risks of having an inconsistent or unphysical set of parameters.
This risk has been mitigated for the pure SSC and SSC+EIC models through parameter constraints, but the final assessment of interpreting the quality and relevance of \texttt{Bjet\_MCMC} results is the responsibility of the user.

\section{Conclusion}

\texttt{Bjet\_MCMC} is a new tool in the growing family of SSC models of blazars. Its full version includes 2 SSC zones + EIC, based on \cite{Hervet_2015}. However, only the one-zone pure SSC and one-zone SSC+EIC are fully implemented in the automatic MCMC fitting method. 
\texttt{Bjet\_MCMC} is aimed to be a user-friendly tool that only requires minimal input from the user, namely a configuration file and a SED data file. It is fully parallelized and can take advantage of computing clusters.
The code works as intended and produces consistent results. 
There are other publicly available SSC models at this time that contain SED fitting algorithms. Among the most known are AGNpy \citep{Nigro_2022}, JetSet \citep{Tramacere_2011}, and Naima \citep{naima_2015}. We do not provide any comparison between \texttt{Bjet\_MCMC} and these models in terms of consistency of results, performances, and capabilities. This will be addressed in further studies. However, it has to be noted that \texttt{Bjet\_MCMC} appears to be at the current time the only public tool with an automatic fitting that can handle a full SSC+EIC model with up to 13 free parameters. It needs to be noted that excellent broadband energy coverage of the SED has to be built in order to obtain a good convergence of all parameters for the most extensive model version. This tool needs the user to have some knowledge of the blazar emission processes to understand the limits of SSC models and infer scientific interpretation from their outputs.
Finally, we highlight that the model is still having frequent updates, and some information mentioned in this paper may be quickly outdated. One of our priorities is to explore routes to reduce further the computation time, such as reducing I/O and further optimizing the energy discretization of the radiative components.
The most updated version with relevant information is available publicly on Github.\footnote{\url{https://github.com/Ohervet/Bjet_MCMC}}

\begin{acknowledgments}
We thank David Williams for his advice through the multiple years of development of \texttt{Bjet\_MCMC}. We thank the PHE team members of the LUTH (Paris Observatory) who actively contributed to the early developments of \texttt{Bjet}. We are grateful to the multiple preliminary users of \texttt{Bjet\_MCMC} who provided feedback to improve the code. This work was made possible thanks to the NSF support under grant PHY-2011420.
\end{acknowledgments}

%

\vspace{5mm}


\software{emcee \citep{Foreman_2013}, Astropy \citep{Astropy_2013}, SciPy \citep{Scipy_2001}, NumPy \citep{Numpy_2011}, Matplotlib \citep{Matplotlib_2007})}

\bibliography{Bjet_MCMC.bib}

\begin{thebibliography}{}
\expandafter\ifx\csname natexlab\endcsname\relax\def\natexlab#1{#1}\fi
\providecommand{\url}[1]{\href{#1}{#1}}
\providecommand{\dodoi}[1]{doi:~\href{http://doi.org/#1}{\nolinkurl{#1}}}
\providecommand{\doeprint}[1]{\href{http://ascl.net/#1}{\nolinkurl{http://ascl.net/#1}}}
\providecommand{\doarXiv}[1]{\href{https://arxiv.org/abs/#1}{\nolinkurl{https://arxiv.org/abs/#1}}}

\bibitem[{{Adams} {et~al.}(2022){Adams}, {Batshoun}, {Benbow}, {Brill},
  {Buckley}, {Capasso}, {Cavins}, {Christiansen}, {Coppi}, {Errando},
  {Farrell}, {Feng}, {Finley}, {Foote}, {Fortson}, {Furniss}, {Gent}, {Giuri},
  {Hanna}, {Hassan}, {Hervet}, {Holder}, {Houck}, {Humensky}, {Jin}, {Kaaret},
  {Kertzman}, {Kieda}, {Krennrich}, {Kumar}, {Lundy}, {Maier}, {McGrath},
  {Moriarty}, {Mukherjee}, {Nieto}, {Nievas-Rosillo}, {O'Brien}, {Ong},
  {Oppenheimer}, {Otte}, {Patel}, {Pfrang}, {Pohl}, {Prado}, {Pueschel},
  {Quinn}, {Ragan}, {Reynolds}, {Rhatigan}, {Ribeiro}, {Roache}, {Ryan},
  {Santander}, {Sembroski}, {Williams}, {Williamson}, {Valverde}, {Horan},
  {Buson}, {Cheung}, {Ciprini}, {Gasparrini}, {Ojha}, {van Zyl}, \&
  {Sironi}}]{Adams_2022}
{Adams}, C.~B., {Batshoun}, J., {Benbow}, W., {et~al.} 2022, \apj, 924, 95,
  \dodoi{10.3847/1538-4357/ac32bd}

\bibitem[{{Archer} {et~al.}(2018){Archer}, {Benbow}, {Bird}, {Brose},
  {Buchovecky}, {Bugaev}, {Cui}, {Daniel}, {Falcone}, {Feng}, {Finley},
  {Flinders}, {Fortson}, {Furniss}, {Gillanders}, {H{\"u}tten}, {Hanna},
  {Hervet}, {Holder}, {Hughes}, {Humensky}, {Johnson}, {Kaaret}, {Kar},
  {Kelley-Hoskins}, {Kieda}, {Krause}, {Krennrich}, {Kumar}, {Lang}, {Lin},
  {McArthur}, {Moriarty}, {Mukherjee}, {Nieto}, {O'Brien}, {Ong}, {Otte},
  {Park}, {Petrashyk}, {Pohl}, {Popkow}, {Pueschel}, {Quinn}, {Ragan},
  {Reynolds}, {Richards}, {Roache}, {Rulten}, {Sadeh}, {Sembroski},
  {Shahinyan}, {Tyler}, {Wakely}, {Weiner}, {Weinstein}, {Wells}, {Wilcox},
  {Wilhelm}, {Williams}, {VERITAS Collaboration}, {Brisken}, \&
  {Pontrelli}}]{Archer_2018}
{Archer}, A., {Benbow}, W., {Bird}, R., {et~al.} 2018, \apj, 862, 41,
  \dodoi{10.3847/1538-4357/aacbd0}

\bibitem[{{Astropy Collaboration} {et~al.}(2013){Astropy Collaboration},
  {Robitaille}, {Tollerud}, {Greenfield}, {Droettboom}, {Bray}, {Aldcroft},
  {Davis}, {Ginsburg}, {Price-Whelan}, {Kerzendorf}, {Conley}, {Crighton},
  {Barbary}, {Muna}, {Ferguson}, {Grollier}, {Parikh}, {Nair}, {Unther},
  {Deil}, {Woillez}, {Conseil}, {Kramer}, {Turner}, {Singer}, {Fox}, {Weaver},
  {Zabalza}, {Edwards}, {Azalee Bostroem}, {Burke}, {Casey}, {Crawford},
  {Dencheva}, {Ely}, {Jenness}, {Labrie}, {Lim}, {Pierfederici}, {Pontzen},
  {Ptak}, {Refsdal}, {Servillat}, \& {Streicher}}]{Astropy_2013}
{Astropy Collaboration}, {Robitaille}, T.~P., {Tollerud}, E.~J., {et~al.} 2013,
  \aap, 558, A33, \dodoi{10.1051/0004-6361/201322068}

\bibitem[{{Avni}(1976)}]{Avni_1976}
{Avni}, Y. 1976, \apj, 210, 642, \dodoi{10.1086/154870}

\bibitem[{{Bennett} {et~al.}(2014){Bennett}, {Larson}, {Weiland}, \&
  {Hinshaw}}]{Bennett_2014}
{Bennett}, C.~L., {Larson}, D., {Weiland}, J.~L., \& {Hinshaw}, G. 2014, \apj,
  794, 135, \dodoi{10.1088/0004-637X/794/2/135}

\bibitem[{{Cerruti} {et~al.}(2013){Cerruti}, {Boisson}, \&
  {Zech}}]{Cerruti_2013}
{Cerruti}, M., {Boisson}, C., \& {Zech}, A. 2013, \aap, 558, A47,
  \dodoi{10.1051/0004-6361/201220963}

\bibitem[{{Foreman-Mackey} {et~al.}(2013){Foreman-Mackey}, {Hogg}, {Lang}, \&
  {Goodman}}]{Foreman_2013}
{Foreman-Mackey}, D., {Hogg}, D.~W., {Lang}, D., \& {Goodman}, J. 2013, \pasp,
  125, 306, \dodoi{10.1086/670067}

\bibitem[{{Franceschini} \& {Rodighiero}(2017)}]{Franceschini_2017}
{Franceschini}, A., \& {Rodighiero}, G. 2017, \aap, 603, A34,
  \dodoi{10.1051/0004-6361/201629684}

\bibitem[{{Ghisellini} \& {Madau}(1996)}]{Ghisellini_1996}
{Ghisellini}, G., \& {Madau}, P. 1996, \mnras, 280, 67,
  \dodoi{10.1093/mnras/280.1.67}

\bibitem[{{Ghisellini} {et~al.}(1985){Ghisellini}, {Maraschi}, \&
  {Treves}}]{Ghisellini_1985}
{Ghisellini}, G., {Maraschi}, L., \& {Treves}, A. 1985, \aap, 146, 204

\bibitem[{{Ghisellini} {et~al.}(2005){Ghisellini}, {Tavecchio}, \&
  {Chiaberge}}]{Ghisellini_2005}
{Ghisellini}, G., {Tavecchio}, F., \& {Chiaberge}, M. 2005, \aap, 432, 401,
  \dodoi{10.1051/0004-6361:20041404}

\bibitem[{{Ginzburg} \& {Syrovatskii}(1965)}]{Ginzburg_1965}
{Ginzburg}, V.~L., \& {Syrovatskii}, S.~I. 1965, \araa, 3, 297,
  \dodoi{10.1146/annurev.aa.03.090165.001501}

\bibitem[{{Ginzburg} \& {Syrovatskii}(1969)}]{Ginzburg_1969}
---. 1969, \araa, 7, 375, \dodoi{10.1146/annurev.aa.07.090169.002111}

\bibitem[{{Goodman} \& {Weare}(2010)}]{Goodman_2010}
{Goodman}, J., \& {Weare}, J. 2010, Communications in Applied Mathematics and
  Computational Science, 5, 65, \dodoi{10.2140/camcos.2010.5.65}

\bibitem[{{Gould}(1979)}]{Gould_1979}
{Gould}, R.~J. 1979, \aap, 76, 306

\bibitem[{Hervet(2015)}]{Hervet_these_2015}
Hervet, O. 2015, Theses, {Observatoire de paris}.
\newblock \url{https://hal.science/tel-01240215}

\bibitem[{{Hervet} {et~al.}(2015){Hervet}, {Boisson}, \& {Sol}}]{Hervet_2015}
{Hervet}, O., {Boisson}, C., \& {Sol}, H. 2015, \aap, 578, A69,
  \dodoi{10.1051/0004-6361/201425330}

\bibitem[{{Hervet} {et~al.}(2016){Hervet}, {Boisson}, \& {Sol}}]{Hervet_2016}
---. 2016, \aap, 592, A22, \dodoi{10.1051/0004-6361/201628117}

\bibitem[{{H.E.S.S. Collaboration} {et~al.}(2012){H.E.S.S. Collaboration},
  {Abramowski, A.}, {Acero, F.}, {Aharonian, F.}, {Akhperjanian, A. G.},
  {Anton, G.}, {Balzer, A.}, {Barnacka, A.}, {Becherini, Y.}, {Becker, J.},
  {Bernl\"ohr, K.}, {Birsin, E.}, {Biteau, J.}, {Bochow, A.}, {Boisson, C.},
  {Bolmont, J.}, {Bordas, P.}, {Brucker, J.}, {Brun, F.}, {Brun, P.}, {Bulik,
  T.}, {B\"usching, I.}, {Carrigan, S.}, {Casanova, S.}, {Cerruti, M.},
  {Chadwick, P. M.}, {Charbonnier, A.}, {Chaves, R. C. G.}, {Cheesebrough, A.},
  {Cologna, G.}, {Conrad, J.}, {Dalton, M.}, {Daniel, M. K.}, {Davids, I. D.},
  {Degrange, B.}, {Deil, C.}, {Dickinson, H. J.}, {Djannati-Ata\"{\i}, A.},
  {Domainko, W.}, {Drury, L. O\'{}C.}, {Dubus, G.}, {Dutson, K.}, {Dyks, J.},
  {Dyrda, M.}, {Egberts, K.}, {Eger, P.}, {Espigat, P.}, {Fallon, L.}, {Fegan,
  S.}, {Feinstein, F.}, {Fernandes, M. V.}, {Fiasson, A.}, {Fontaine, G.},
  {F\"orster, A.}, {F\"u\ss{}ling, M.}, {Gallant, Y. A.}, {Gast, H.},
  {G\'erard, L.}, {Gerbig, D.}, {Giebels, B.}, {Glicenstein, J. F.}, {Gl\"uck,
  B.}, {G\"oring, D.}, {H\"affner, S.}, {Hague, J. D.}, {Hahn, J.}, {Hampf,
  D.}, {Harris, J.}, {Hauser, M.}, {Heinz, S.}, {Heinzelmann, G.}, {Henri, G.},
  {Hermann, G.}, {Hillert, A.}, {Hinton, J. A.}, {Hofmann, W.}, {Hofverberg,
  P.}, {Holler, M.}, {Horns, D.}, {Jacholkowska, A.}, {de Jager, O. C.}, {Jahn,
  C.}, {Jamrozy, M.}, {Jung, I.}, {Kastendieck, M. A.}, {Katarzy\'{}nski, K.},
  {Katz, U.}, {Kaufmann, S.}, {Keogh, D.}, {Kh\'elifi, B.}, {Klochkov, D.},
  {Klu\'{}zniak, W.}, {Kneiske, T.}, {Komin, Nu.}, {Kosack, K.}, {Kossakowski,
  R.}, {Krayzel, F.}, {Laffon, H.}, {Lamanna, G.}, {Lenain, J.-P.}, {Lennarz,
  D.}, {Lohse, T.}, {Lopatin, A.}, {Lu, C.-C.}, {Marandon, V.}, {Marcowith,
  A.}, {Masbou, J.}, {Maxted, N.}, {Mayer, M.}, {McComb, T. J. L.}, {Medina, M.
  C.}, {M\'ehault, J.}, {Moderski, R.}, {Mohamed, M.}, {Moulin, E.}, {Naumann,
  C. L.}, {Naumann-Godo, M.}, {de Naurois, M.}, {Nedbal, D.}, {Nekrassov, D.},
  {Nguyen, N.}, {Nicholas, B.}, {Niemiec, J.}, {Nolan, S. J.}, {Ohm, S.}, {de
  O\~na Wilhelmi, E.}, {Opitz, B.}, {Ostrowski, M.}, {Oya, I.}, {Panter, M.},
  {Paz Arribas, M.}, {Pekeur, N. W.}, {Pelletier, G.}, {Perez, J.}, {Petrucci,
  P.-O.}, {Peyaud, B.}, {Pita, S.}, {P\"uhlhofer, G.}, {Punch, M.},
  {Quirrenbach, A.}, {Raue, M.}, {Rayner, S. M.}, {Reimer, A.}, {Reimer, O.},
  {Renaud, M.}, {de los Reyes, R.}, {Rieger, F.}, {Ripken, J.}, {Rob, L.},
  {Rosier-Lees, S.}, {Rowell, G.}, {Rudak, B.}, {Rulten, C. B.}, {Sahakian,
  V.}, {Sanchez, D. A.}, {Santangelo, A.}, {Schlickeiser, R.}, {Schulz, A.},
  {Schwanke, U.}, {Schwarzburg, S.}, {Schwemmer, S.}, {Sheidaei, F.}, {Skilton,
  J. L.}, {Sol, H.}, {Spengler, G.}, {Stawarz, L.}, {Steenkamp, R.}, {Stegmann,
  C.}, {Stinzing, F.}, {Stycz, K.}, {Sushch, I.}, {Szostek, A.}, {Tavernet,
  J.-P.}, {Terrier, R.}, {Tluczykont, M.}, {Valerius, K.}, {van Eldik, C.},
  {Vasileiadis, G.}, {Venter, C.}, {Viana, A.}, {Vincent, P.}, {V\"olk, H. J.},
  {Volpe, F.}, {Vorobiov, S.}, {Vorster, M.}, {Wagner, S. J.}, {Ward, M.},
  {White, R.}, {Wierzcholska, A.}, {Zacharias, M.}, {Zajczyk, A.}, {Zdziarski,
  A. A.}, {Zech, A.}, \& {Zechlin, H.-S.}}]{HESS_2012}
{H.E.S.S. Collaboration}, {Abramowski, A.}, {Acero, F.}, {et~al.} 2012, A\&A,
  542, A94, \dodoi{10.1051/0004-6361/201218910}

\bibitem[{{HESS Collaboration} {et~al.}(2018){HESS Collaboration}, {Abdalla},
  {Abramowski}, {Aharonian}, {Ait Benkhali}, {Akhperjanian}, {Andersson},
  {Ang{\"u}ner}, {Arrieta}, {Aubert}, {Backes}, {Balzer}, {Barnard},
  {Becherini}, {Becker Tjus}, {Berge}, {Bernhard}, {Bernl{\"o}hr}, {Blackwell},
  {B{\"o}ttcher}, {Boisson}, {Bolmont}, {Bordas}, {Bregeon}, {Brun}, {Brun},
  {Bryan}, {Bulik}, {Capasso}, {Carr}, {Casanova}, {Cerruti}, {Chakraborty},
  {Chalme-Calvet}, {Chaves}, {Chen}, {Chevalier}, {Chr{\'e}tien},
  {Colafrancesco}, {Cologna}, {Condon}, {Conrad}, {Cui}, {Davids}, {Decock},
  {Degrange}, {Deil}, {Devin}, {deWilt}, {Dirson}, {Djannati-Ata{\"\i}},
  {Domainko}, {Donath}, {Drury}, {Dubus}, {Dutson}, {Dyks}, {Dyrda}, {Edwards},
  {Egberts}, {Eger}, {Ernenwein}, {Eschbach}, {Farnier}, {Fegan}, {Fernandes},
  {Fiasson}, {Fontaine}, {F{\"o}rster}, {Funk}, {F{\"u}{\ss}ling}, {Gabici},
  {Gajdus}, {Gallant}, {Garrigoux}, {Giavitto}, {Giebels}, {Glicenstein},
  {Gottschall}, {Goyal}, {Grondin}, {Hadasch}, {Hahn}, {Haupt}, {Hawkes},
  {Heinzelmann}, {Henri}, {Hermann}, {Hervet}, {Hinton}, {Hofmann}, {Hoischen},
  {Holler}, {Horns}, {Ivascenko}, {Jacholkowska}, {Jamrozy}, {Janiak},
  {Jankowsky}, {Jankowsky}, {Jingo}, {Jogler}, {Jouvin}, {Jung-Richardt},
  {Kastendieck}, {Katarzy{\'n}ski}, {Katz}, {Kerszberg}, {Kh{\'e}lifi},
  {Kieffer}, {King}, {Klepser}, {Klochkov}, {Klu{\'z}niak}, {Kolitzus},
  {Komin}, {Kosack}, {Krakau}, {Kraus}, {Krayzel}, {Kr{\"u}ger}, {Laffon},
  {Lamanna}, {Lau}, {Lees}, {Lefaucheur}, {Lefranc}, {Lemi{\`e}re},
  {Lemoine-Goumard}, {Lenain}, {Leser}, {Lohse}, {Lorentz}, {Liu},
  {L{\'o}pez-Coto}, {Lypova}, {Marandon}, {Marcowith}, {Mariaud}, {Marx},
  {Maurin}, {Maxted}, {Mayer}, {Meintjes}, {Meyer}, {Mitchell}, {Moderski},
  {Mohamed}, {Mohrmann}, {Mor{\^a}}, {Moulin}, {Murach}, {de Naurois},
  {Niederwanger}, {Niemiec}, {Oakes}, {O'Brien}, {Odaka}, {{\"O}ttl}, {Ohm},
  {Ostrowski}, {Oya}, {Padovani}, {Panter}, {Parsons}, {Pekeur}, {Pelletier},
  {Perennes}, {Petrucci}, {Peyaud}, {Piel}, {Pita}, {Poon}, {Prokhorov},
  {Prokoph}, {P{\"u}hlhofer}, {Punch}, {Quirrenbach}, {Raab}, {Reimer},
  {Reimer}, {Renaud}, {de los Reyes}, {Rieger}, {Romoli}, {Rosier-Lees},
  {Rowell}, {Rudak}, {Rulten}, {Sahakian}, {Salek}, {Sanchez}, {Santangelo},
  {Sasaki}, {Schlickeiser}, {Sch{\"u}ssler}, {Schulz}, {Schwanke}, {Schwemmer},
  {Settimo}, {Seyffert}, {Shafi}, {Shilon}, {Simoni}, {Sol}, {Spanier},
  {Spengler}, {Spies}, {Stawarz}, {Steenkamp}, {Stegmann}, {Stinzing}, {Stycz},
  {Sushch}, {Tavernet}, {Tavernier}, {Taylor}, {Terrier}, {Tibaldo}, {Tiziani},
  {Tluczykont}, {Trichard}, {Tuffs}, {Uchiyama}, {van der Walt}, {van Eldik},
  {van Rensburg}, {van Soelen}, {Vasileiadis}, {Veh}, {Venter}, {Viana},
  {Vincent}, {Vink}, {Voisin}, {V{\"o}lk}, {Vuillaume}, {Wadiasingh}, {Wagner},
  {Wagner}, {Wagner}, {White}, {Wierzcholska}, {Willmann}, {W{\"o}rnlein},
  {Wouters}, {Yang}, {Zabalza}, {Zaborov}, {Zacharias}, {Zanin}, {Zdziarski},
  {Zech}, {Zefi}, {Ziegler}, \& {{\.Z}ywucka}}]{HESS_2018}
{HESS Collaboration}, {Abdalla}, H., {Abramowski}, A., {et~al.} 2018, \mnras,
  476, 4187, \dodoi{10.1093/mnras/sty439}

\bibitem[{Hunter(2007)}]{Matplotlib_2007}
Hunter, J.~D. 2007, Computing in Science \& Engineering, 9, 90,
  \dodoi{10.1109/MCSE.2007.55}

\bibitem[{{Inoue} \& {Takahara}(1996)}]{Inoue_1996}
{Inoue}, S., \& {Takahara}, F. 1996, \apj, 463, 555, \dodoi{10.1086/177270}

\bibitem[{{Jim{\'e}nez-Fern{\'a}ndez} \& {van Eerten}(2021)}]{Jim_2021}
{Jim{\'e}nez-Fern{\'a}ndez}, B., \& {van Eerten}, H.~J. 2021, \mnras, 500,
  3613, \dodoi{10.1093/mnras/staa3163}

\bibitem[{Jones {et~al.}(2001--)Jones, Oliphant, Peterson,
  {et~al.}}]{Scipy_2001}
Jones, E., Oliphant, T., Peterson, P., {et~al.} 2001--, {SciPy}: Open source
  scientific tools for {Python}.
\newblock \url{http://www.scipy.org/}

\bibitem[{{Katarzy{\'n}ski} {et~al.}(2001){Katarzy{\'n}ski}, {Sol}, \&
  {Kus}}]{Katarzynski_2001}
{Katarzy{\'n}ski}, K., {Sol}, H., \& {Kus}, A. 2001, \aap, 367, 809,
  \dodoi{10.1051/0004-6361:20000538}

\bibitem[{{Lampton} {et~al.}(1976){Lampton}, {Margon}, \&
  {Bowyer}}]{Lampton_1976}
{Lampton}, M., {Margon}, B., \& {Bowyer}, S. 1976, \apj, 208, 177,
  \dodoi{10.1086/154592}

\bibitem[{{Longair}(1994)}]{Longair_1994}
{Longair}, M.~S. 1994, {High energy astrophysics. Vol.2: Stars, the galaxy and
  the interstellar medium}, Vol.~2

\bibitem[{MacKay(2003)}]{MacKay_2003}
MacKay, D. J.~C. 2003, Information Theory, Inference, and Learning Algorithms
  (Cambridge University Press).
\newblock \url{http://www.cambridge.org/0521642981}

\bibitem[{{Maraschi} \& {Rovetti}(1994)}]{Maraschi_1994}
{Maraschi}, L., \& {Rovetti}, F. 1994, \apj, 436, 79, \dodoi{10.1086/174882}

\bibitem[{{Marscher}(1980)}]{Marscher_1980}
{Marscher}, A.~P. 1980, \apj, 235, 386, \dodoi{10.1086/157642}

\bibitem[{{Marscher} \& {Gear}(1985)}]{Marscher_1985}
{Marscher}, A.~P., \& {Gear}, W.~K. 1985, \apj, 298, 114,
  \dodoi{10.1086/163592}

\bibitem[{{Nalewajko} {et~al.}(2014){Nalewajko}, {Begelman}, \&
  {Sikora}}]{Nalewajko_2014}
{Nalewajko}, K., {Begelman}, M.~C., \& {Sikora}, M. 2014, \apj, 789, 161,
  \dodoi{10.1088/0004-637X/789/2/161}

\bibitem[{{Nigro} {et~al.}(2022){Nigro}, {Sitarek}, {Gliwny}, {Sanchez},
  {Tramacere}, \& {Craig}}]{Nigro_2022}
{Nigro}, C., {Sitarek}, J., {Gliwny}, P., {et~al.} 2022, \aap, 660, A18,
  \dodoi{10.1051/0004-6361/202142000}

\bibitem[{{Padovani} \& {Giommi}(1995)}]{Padovani_1995}
{Padovani}, P., \& {Giommi}, P. 1995, \apj, 444, 567, \dodoi{10.1086/175631}

\bibitem[{{Qin} {et~al.}(2018){Qin}, {Wang}, {Yang}, {Yuan}, {Mao}, \&
  {Kang}}]{Qin_2018}
{Qin}, L., {Wang}, J., {Yang}, C., {et~al.} 2018, \pasj, 70, 5,
  \dodoi{10.1093/pasj/psx150}

\bibitem[{{Sikora} {et~al.}(1994){Sikora}, {Begelman}, \& {Rees}}]{Sikora_1994}
{Sikora}, M., {Begelman}, M.~C., \& {Rees}, M.~J. 1994, \apj, 421, 153,
  \dodoi{10.1086/173633}

\bibitem[{{Tavecchio} {et~al.}(2011){Tavecchio}, {Becerra-Gonzalez},
  {Ghisellini}, {Stamerra}, {Bonnoli}, {Foschini}, \&
  {Maraschi}}]{Tavecchio_2011}
{Tavecchio}, F., {Becerra-Gonzalez}, J., {Ghisellini}, G., {et~al.} 2011, \aap,
  534, A86, \dodoi{10.1051/0004-6361/201117204}

\bibitem[{{Tavecchio} {et~al.}(1998){Tavecchio}, {Maraschi}, \&
  {Ghisellini}}]{Tavecchio_1998}
{Tavecchio}, F., {Maraschi}, L., \& {Ghisellini}, G. 1998, \apj, 509, 608,
  \dodoi{10.1086/306526}

\bibitem[{{Tramacere} {et~al.}(2011){Tramacere}, {Massaro}, \&
  {Taylor}}]{Tramacere_2011}
{Tramacere}, A., {Massaro}, E., \& {Taylor}, A.~M. 2011, \apj, 739, 66,
  \dodoi{10.1088/0004-637X/739/2/66}

\bibitem[{{Urry} \& {Padovani}(1995)}]{Urry_1995}
{Urry}, C.~M., \& {Padovani}, P. 1995, \pasp, 107, 803, \dodoi{10.1086/133630}

\bibitem[{{Valverde} {et~al.}(2020){Valverde}, {Horan}, {Bernard}, {Fegan},
  {Fermi-LAT Collaboration}, {Abeysekara}, {Archer}, {Benbow}, {Bird}, {Brill},
  {Brose}, {Buchovecky}, {Buckley}, {Christiansen}, {Cui}, {Falcone}, {Feng},
  {Finley}, {Fortson}, {Furniss}, {Gent}, {Gillanders}, {Giuri}, {Gueta},
  {Hanna}, {Hassan}, {Hervet}, {Holder}, {Hughes}, {Humensky}, {Kaaret},
  {Kelley-Hoskins}, {Kertzman}, {Kieda}, {Krause}, {Krennrich}, {Lang},
  {Maier}, {Moriarty}, {Mukherjee}, {Nieto}, {Nievas-Rosillo}, {O'Brien},
  {Ong}, {Otte}, {Park}, {Petrashyk}, {Pfrang}, {Pichel}, {Pohl}, {Prado},
  {Pueschel}, {Quinn}, {Ragan}, {Reynolds}, {Ribeiro}, {Richards}, {Roache},
  {Sadeh}, {Santander}, {Scott}, {Sembroski}, {Shahinyan}, {Shang}, {Sushch},
  {Vassiliev}, {Weinstein}, {Wells}, {Wilcox}, {Wilhelm}, {Williams},
  {Williamson}, {VERITAS Collaboration}, {Noto}, {Edwards}, {Piner}, {Fallah
  Ramazani}, {Hovatta}, {Jormanainen}, {Lindfors}, {Nilsson}, {Takalo},
  {Kovalev}, {Lister}, {Pushkarev}, {Savolainen}, {Kiehlmann}, {Max-Moerbeck},
  {Readhead}, {L{\"a}hteenm{\"a}ki}, \& {Tornikoski}}]{Valverde_2020}
{Valverde}, J., {Horan}, D., {Bernard}, D., {et~al.} 2020, \apj, 891, 170,
  \dodoi{10.3847/1538-4357/ab765d}

\bibitem[{Walt {et~al.}(2011)Walt, Colbert, \& Varoquaux}]{Numpy_2011}
Walt, S. v.~d., Colbert, S.~C., \& Varoquaux, G. 2011, Computing in Science \&
  Engineering, 13, 22, \dodoi{10.1109/MCSE.2011.37}

\bibitem[{{Zabalza}(2015)}]{naima_2015}
{Zabalza}, V. 2015, Proc.~of International Cosmic Ray Conference 2015, 922

\end{thebibliography}
\bibliographystyle{aasjournal}

\appendix
\section{Additional \texttt{Bjet\_MCMC} outputs for 1RXS J101015.9-311900 }

\begin{figure}[h!]
\centering
\includegraphics[width=.49\textwidth]{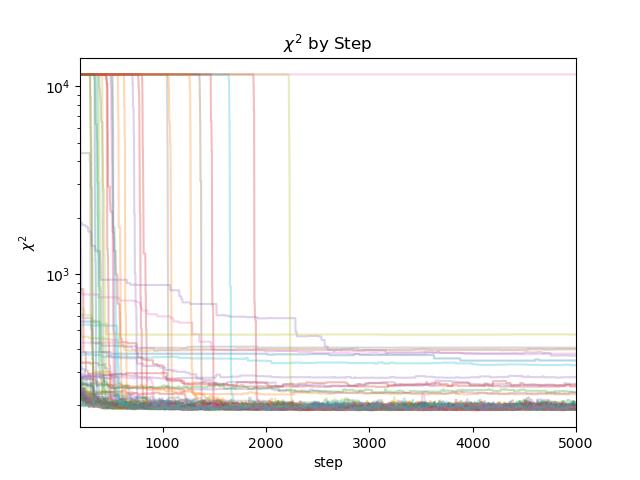}
\includegraphics[width=.49\textwidth]{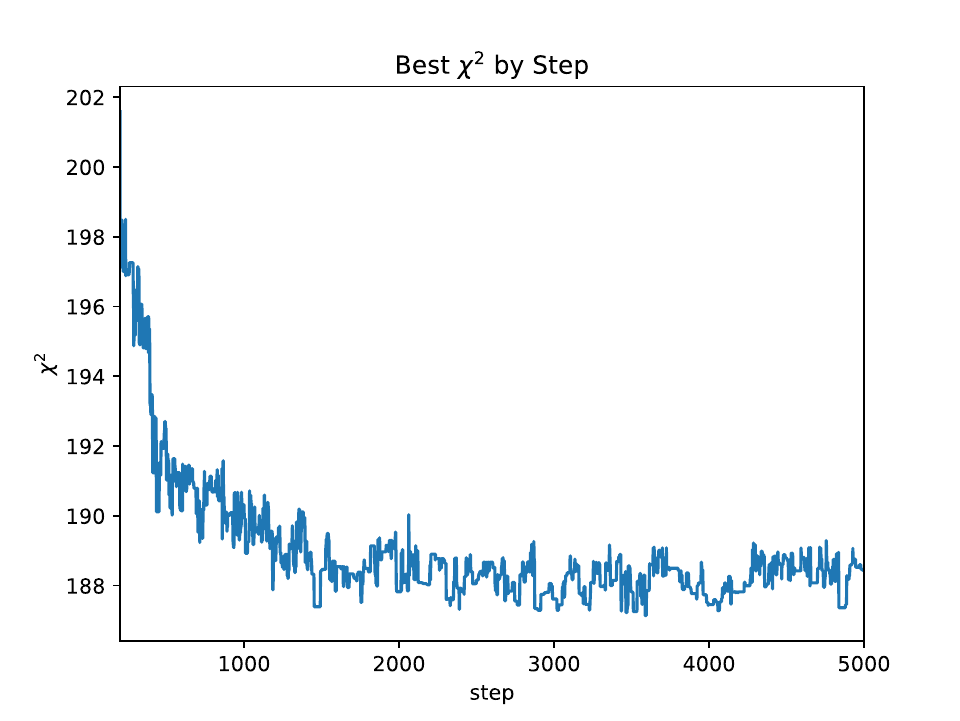}
\includegraphics[width=.49\textwidth]{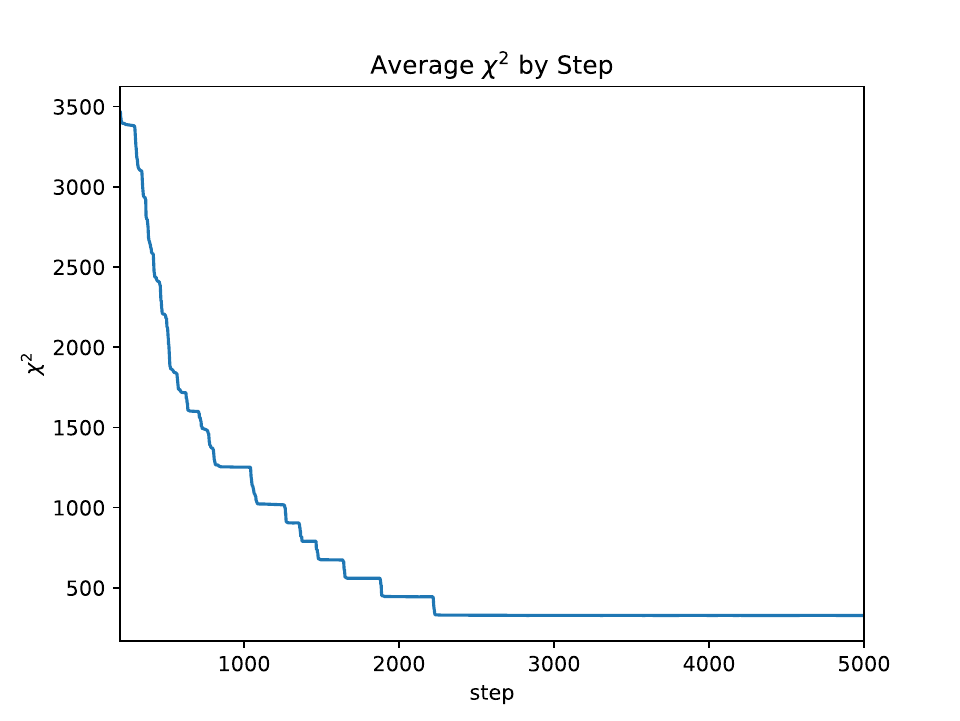}
\caption{\texttt{Bjet\_MCMC} $\chi^2$ values of walkers at each step for the SED fit of 1RXS J101015.9-311900. \textit{Upper-left}: all walkers,  \textit{Upper-right}: smallest $\chi^2$, \textit{Bottom}: average $\chi^2$.}
\label{Fig::Chi2_1RXS}
\end{figure}

\begin{figure}[h!]
\centering
\includegraphics[width=0.8\textwidth]{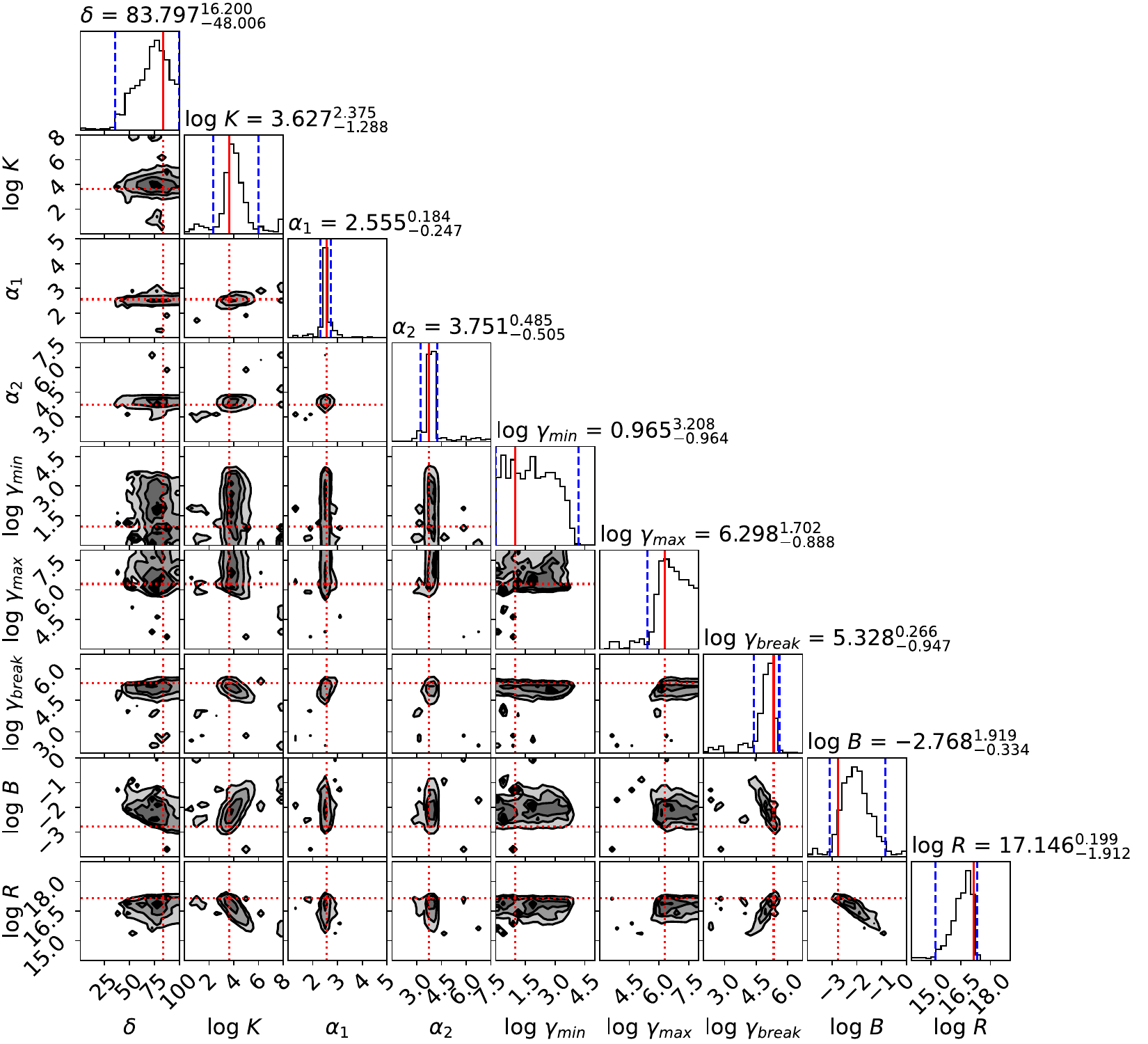}
\caption{Corner plot of the posterior probability distribution of the free parameters from the SED fit of 1RXS J101015.9-311900.}
\end{figure}

\newpage
\section{Additional \texttt{Bjet\_MCMC} outputs for PKS 1222+216 }

\begin{figure}[h!]
\centering
\includegraphics[width=.49\textwidth]{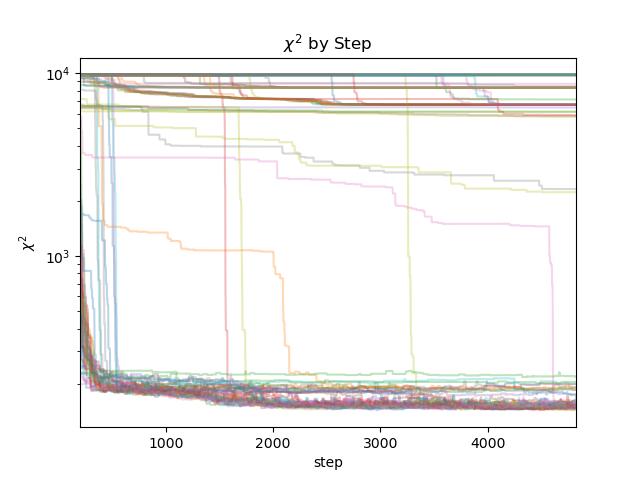}
\includegraphics[width=.49\textwidth]{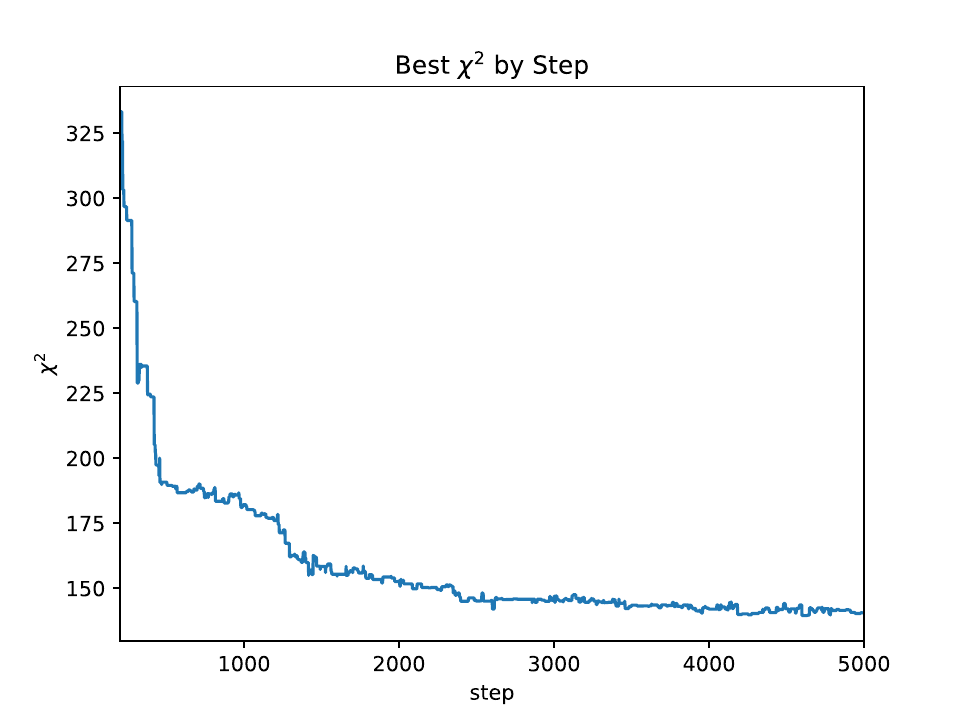}
\includegraphics[width=.49\textwidth]{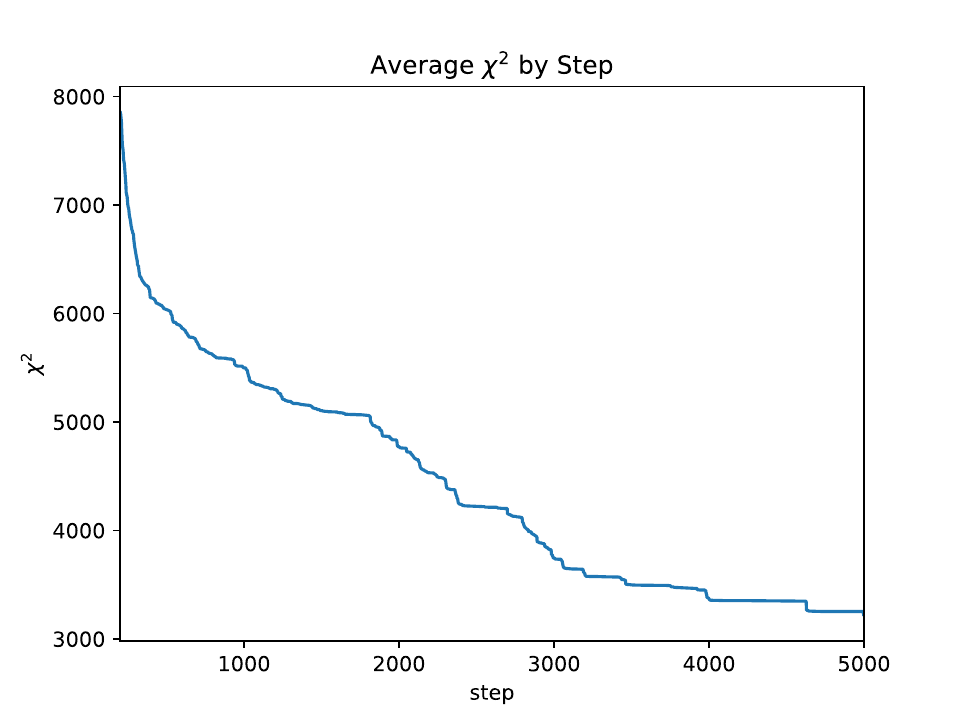}
\caption{\texttt{Bjet\_MCMC} $\chi^2$ values of walkers at each step for the SED fit of PKS 1222+216. \textit{Upper-left}: all walkers,  \textit{Upper-right}: smallest $\chi^2$, \textit{Bottom}: average $\chi^2$.}
\label{Fig::Chi2_PKS1222}
\end{figure}

\begin{figure}[h!]
\centering
\includegraphics[width=\textwidth]{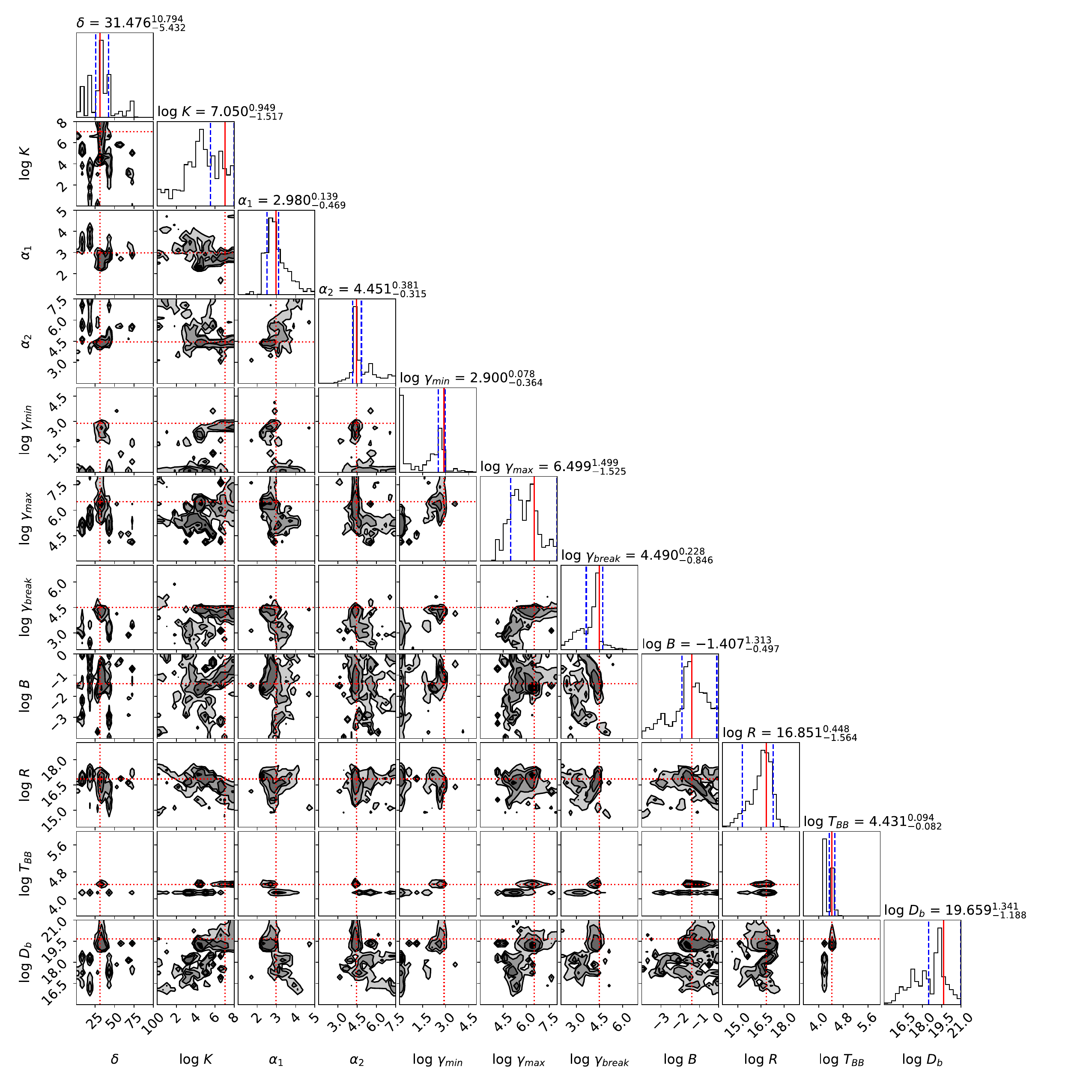}
\caption{Corner plot of the posterior probability distribution of the free parameters from the SED fit of PKS 1222+216.}
\label{Fig::Corner_PKS1222}
\end{figure}





\end{document}